\documentclass[%
notitlepage,
reprint,
superscriptaddress,
amsmath,amssymb,
aps,
]{revtex4-2}

\usepackage{xcolor}
\usepackage{lipsum}
\usepackage{graphicx}
\usepackage{dcolumn}
\usepackage{bm}
\usepackage{hyperref}

\usepackage{epstopdf}

\newcommand{\bra}[1]{\langle #1 \vert}
\newcommand{\ket}[1]{\vert #1 \rangle}

\usepackage{ulem}

\begin{document}
	
	
	\title{Managing rogue quantum amplitudes: a control perspective in quantum walks}
	
	\affiliation{Laborat\' orio de F\'isica Te\' orica e Computacional, Departamento de F\'isica, Universidade Federal de Pernambuco, 50670-901 Recife, PE, Brazil.}

	\author{A. R. C. Buarque$^{1}$}
	\author{E. P. Raposo$^{1}$}

\begin{abstract}
    We investigate the emergence of rogue quantum amplitudes in discrete-time quantum walks (DTQWs) influenced by phase disorder. 
    Our study reveals the statistics of occupation probability amplitudes in space and time, uncovering optimal disorder regimes that favor rogue wave events. 
    Through numerical simulations, we demonstrate that the probability of rogue waves increases with quantum coins close to the Pauli-Z choice, regardless the disorder degree. 
    Conversely, for coins near Pauli-X rogue events are scarce, except under weak disorder. 
    A monotonic threshold is observed between rare- and high-probability rogue wave regimes, depending on the quantum coin. 
    We provide a comprehensive analysis of the coin-disorder interplay to rogue wave events. 
    Our findings shed light on the possible control of extreme quantum amplitudes through quantum coins in disordered~DTQWs.
\end{abstract}
	\maketitle
	
\section{Introduction}

Understanding the emergence of unlikely extreme events in nature has long been a topic of great interest~\cite{bo1,bo2,bo3}, with applications in fields as diverse as epidemics~\cite{epi}, photonics~\cite{pho}, and neurobiology~\cite{neu}, to name a few. 
One of such phenomena is known as rogue waves (RWs)~\cite{kharif2008rogue, haver2004possible}. 
In the maritime community, RWs have been described as waves with amplitudes far exceeding what is expected for the prevailing sea state. 
In this context, RWs represent unpredictable oceanic waves with large amplitudes that seemingly materialize out of nowhere and vanish without~a~trace. 

The intriguing and often unpredictable occurrence of~RWs has captivated the attention of researchers across various scientific domains, from oceanography~\cite{haver2004possible} to optics~\cite{solli2007optical} and many others (for comprehensive reviews, see, e.g,  Refs.~\cite{dudley2019rogue,dudley2014instabilities}). 
In particular, the connection between the oceanic RW phenomenon and light propagation within optical fibers has gained prominence, specially in the framework of the nonlinear Schrödinger equation~\cite{solli2007optical}. 
This link has spurred a surge of interest in wave phenomena exhibiting long-tailed statistical distributions, whose associated outlier events greatly surpass the predictions from Gaussian statistics. 
RWs have also been extensively explored in diverse other contexts, including linear and nonlinear optics \cite{PhysRevE.102.052219,PhysRevE.84.016604}, plasmas \cite{moslem2011surface}, Bose-Einstein condensates \cite{wen2011matter}, and even finance \cite{yan2011vector}.
From the perspective of nonlinear dynamics related to the emergence of RWs, factors contributing to their occurrence include delayed feedback systems \cite{dal2013extreme,odent2010experimental}, chaotic dynamics in low-dimensional systems \cite{PhysRevLett.107.053901}, soliton collisions \cite{toenger2015emergent}, space-time chaos \cite{PhysRevLett.116.013901}, vortex turbulence \cite{PhysRevLett.116.043903}, and integrable turbulence \cite{PhysRevLett.114.143903}. 

One of the central issues in the study of RWs lies in the understanding of how these events arise, an objective closely related to the will of predicting and controlling extreme events. 
This challenge stems from the multifaceted processes involved in the RW~formation. 
Extensive debates persist regarding whether RWs originate from linear \cite{PhysRevLett.104.093901} or nonlinear \cite{ying2011linear,PhysRevE.80.026601,tlidi2022rogue} processes, and the role of disorder in this narrative \cite{RevModPhys.80.1355,rivas2020rogue}. 
In this context, nonlinear phenomena may further amplify the effects of extreme events that naturally arise from purely linear processes, in a way possibly related to modulational~instability~\cite{dudley2014instabilities,bezerra2023thresholds}. 
RWs have also been explored in the quantum mechanical context. 
The emergence of RWs in quantum chains has introduced novel aspects to the understanding of a number of quantum dynamical regimes, particularly in disordered media \cite{RevModPhys.80.1355,rivas2020rogue}.
In a previous work~\cite{PhysRevA.107.012425}, we investigated the role of randomness in the formation of anomalous amplitudes in the quantum wave function of a one-dimensional system described by a tight-binding Hamiltonian with correlated on-site disorder. 
We found that a specific effective degree of correlation is responsible for inducing the occurrence of much larger extreme amplitude events, particularly when compared to the case of uncorrelated disorder. 
We remark that similar phenomena  involving correlation have been also  reported in optical systems~\cite{PhysRevE.102.052219}, with the identification of super RWs.

Discrete-time quantum walks (DTQWs) have emerged as a powerful framework for modeling various physical systems and phenomena \cite{wang2013physical}. 
DTQWs have also been recently employed to investigate RW events in the form of sudden, highly-localized extreme wave amplitudes.
In Ref.~\cite{PhysRevA.106.012414} the authors introduced the first model utilizing the approach of DTQWs to study RWs. 
By employing the Hadamard quantum walk with phase disorder, they observed the emergence of RWs in a purely linear system. 
This work also demonstrated that the competition between mobility and localization properties in an intermediate disordered regime is more conducive to the occurrence of extreme events~\cite{PhysRevA.106.012414}. 

However, some fundamental questions arise within the context of DTQWs that remain not addressed so~far: what is the influence of applying different quantum coins (apart from the Hadamard one~\cite{PhysRevA.106.012414}) to the emergence of~RWs? 
Also, can the occurrence of rogue quantum amplitudes be controlled? 
Previous works have shown how various quantum coins can alter fundamental properties of DTQWs, such as transport features in nonlinear \cite{PhysRevA.101.023802} and aperiodic media \cite{PhysRevE.102.012104}, instability and self-focusing characteristics \cite{PhysRevA.103.042213}, diffusivity~\cite{new1}~and entanglement\cite{Wang:18,PhysRevA.89.042307} properties, to name a few.
Thus, investigating the dynamics of the quantum walker under different quantum coins in chains with phase disorder becomes important to understand the emergence of~RWs in DTQWs.

This work aims at bridging between RWs and DTQWs by examining how the inherent wave-like characteristics of quantum walks can shed light on the emergence and dynamics of RWs. 
We demonstrate how the application of various quantum coins influences the occurrence of RWs within the context of one-dimensional DTQWs driven by random phase fluctuations. 
This investigation unveils the characteristic long-tailed statistical behavior of occupation probability, analogous to light intensity observed in optics \cite{gao2020optical}, across the space-time domain. 
Our findings reveal multiple optimal regimes of disorder that maximize the occurrence of these extreme events. 
The RW phenomenon in DTQWs emerges due to a subtle balance between mobility and localization, in which the localization length significantly impacts the walk dynamics. 
We identify a monotonic threshold between quantum walks characterized by rare occurrence of RWs and those with high occurrence probability, displaying a direct dependence on the quantum coin employed in the system. 
Finally, we comprehensively map the relationship between the quantum coins and degree of disorder through a diagram featuring the regions that maximize the emergence of RW~events.

This article is organized as follows. 
In Section~II we introduce the model and describe the general formalism. 
Results and discussions are presented in Section~III. 
Lastly, final remarks and conclusions are left to Section~IV.

\section{Model and formalism}

We consider a quantum random walker propagating in a one-dimensional phase-disordered chain of $N$~sites, with discrete positions indexed by integers $n \, (=1,2,\dots,N)$. 
The quantum walker is defined in a two-level space constituted by the coin space $\mathcal{H}^\mathcal{C} = \{ [ \ket{\uparrow} = (1,0)^T],$ $[\ket{\downarrow} = (0,1)^T] \}$, in which the superscript denotes the transpose, and the position space $\mathcal{H}^{\mathcal{P}} = \{ \ket{n} \}$. 
The Hilbert space is the tensor product $\mathcal{H} = \mathcal{H}^\mathcal{P} \otimes \mathcal{H}^\mathcal{C}$.
The initial state ($t=0$) of the quantum walker is a superposition of the coin and position states in the form 
\begin{equation}
\ket{\Psi(t)}= \sum_n [a_{n}(t)\ket{\uparrow} + b_{n}(t)\ket{\downarrow}] \otimes \ket{n},
\label{eq:initial-state}
\end{equation}
where $a_{n}(t)$ and $b_{n}(t)$ are the probability amplitudes for the up and down coin states at position $n$, respectively. 
The normalization condition is given by $\sum_{n}P_{n}(t)=\sum_{n}[|a_{n}(t)|^2 + |b_{n}(t)|^2] = 1$.   

The system evolution is obtained through $\ket{\psi(t)}=\hat{U}^{t}\ket{\Psi(0)}$, where the time evolution operator $\hat{U}=\hat{S}\hat{C}\hat{D}$
depends on both internal and spatial degrees of freedom of the walker and describes the simultaneous action of the quantum coin $\hat{C}$, conditional displacement $\hat{S}$, and phase-gain~$\hat{D}$ operators. 
Indeed, to account for the internal degrees of freedom a unitary operator $\hat{C}$, known as quantum coin, is applied, which can be expressed as a SU(2) unitary matrix \cite{PhysRevA.77.032326,PhysRevA.40.1371},  
\begin{eqnarray}
\label{quantum-coin}
    \hat{C}(\theta)&=& \cos\theta\ket{\uparrow}\bra{\uparrow} + \sin\theta\ket{\uparrow}\bra{\downarrow} \nonumber \\ 
    & &+\sin\theta\ket{\downarrow}\bra{\uparrow}-\cos\theta\ket{\downarrow}\bra{\downarrow},
\end{eqnarray}
where the angle $0\leq\theta\leq\pi/2$ drives the spatial bias of the quantum coin. 
For example, in the case of a fair coin, which selects both up and down states with equal probability, the choice $\theta=\pi/4$ is adopted (Hadamard coin). 

On the other hand, in order to describe the $N$-cycle architecture, we add periodic boundary conditions to the conditional displacement operator that moves the walker by one lattice spacing at each unit time,
\begin{eqnarray}
    \label{shift-operator}
    \hat{S}= \sum_{n=1}^{N-1}&\ket{\uparrow}&\bra{\uparrow}\otimes |n+1\rangle\langle n|
    + \sum_{n=2}^{N} \ket{\downarrow}\bra{\downarrow}
    \otimes |n-1\rangle\langle n| \nonumber \\
    &+& \ket{\uparrow}\bra{\downarrow}\otimes \ket{1}\bra{N}
    + \ket{\downarrow}\bra{\uparrow}\otimes \ket{N}\bra{1}.
\end{eqnarray}
In addition, the phase-gain operator, defined as
\begin{eqnarray}
\label{eq:phase-operator}
\hat{D}=\sum_{c}\sum_{n}e^{iF(c,n,t)}\ket{c}\bra{c}\otimes\ket{n}\bra{n},
\end{eqnarray}
also plays a relevant role, with $F(c,n,t)$ representing an arbitrary real-valued function and $c=\{\uparrow,\downarrow\}$. 
Actually, the versatility for the choice of $F(c,n,t)$ allows the generation of different dynamic regimes, such as in the investigation of nonlinear and electric field effects in DTQWs \cite{PhysRevA.75.062333,PhysRevA.103.012222,PhysRevA.103.042213}. 
For $F=0$ and $\theta = \pi/4$ the system exhibits the standard Hadamard quantum walk behavior, with the walker spreading out ballistically. 
However, Anderson localization can also emerge by introducing a static random phase modulation characterized by $F(c,n,t)=F(c,n)=2\pi\nu$, where $\nu$~is a number randomly distributed in the range $[-W,W]$ and $W$ represents the width of the disorder. 
So it becomes important to investigate whether RWs can arise in DTQWs under suitable initial conditions and for proper choices of the noise level embedded in $F(c, n)$ and disorder strength.

\begin{figure}
    \centering
\includegraphics[width=\linewidth]{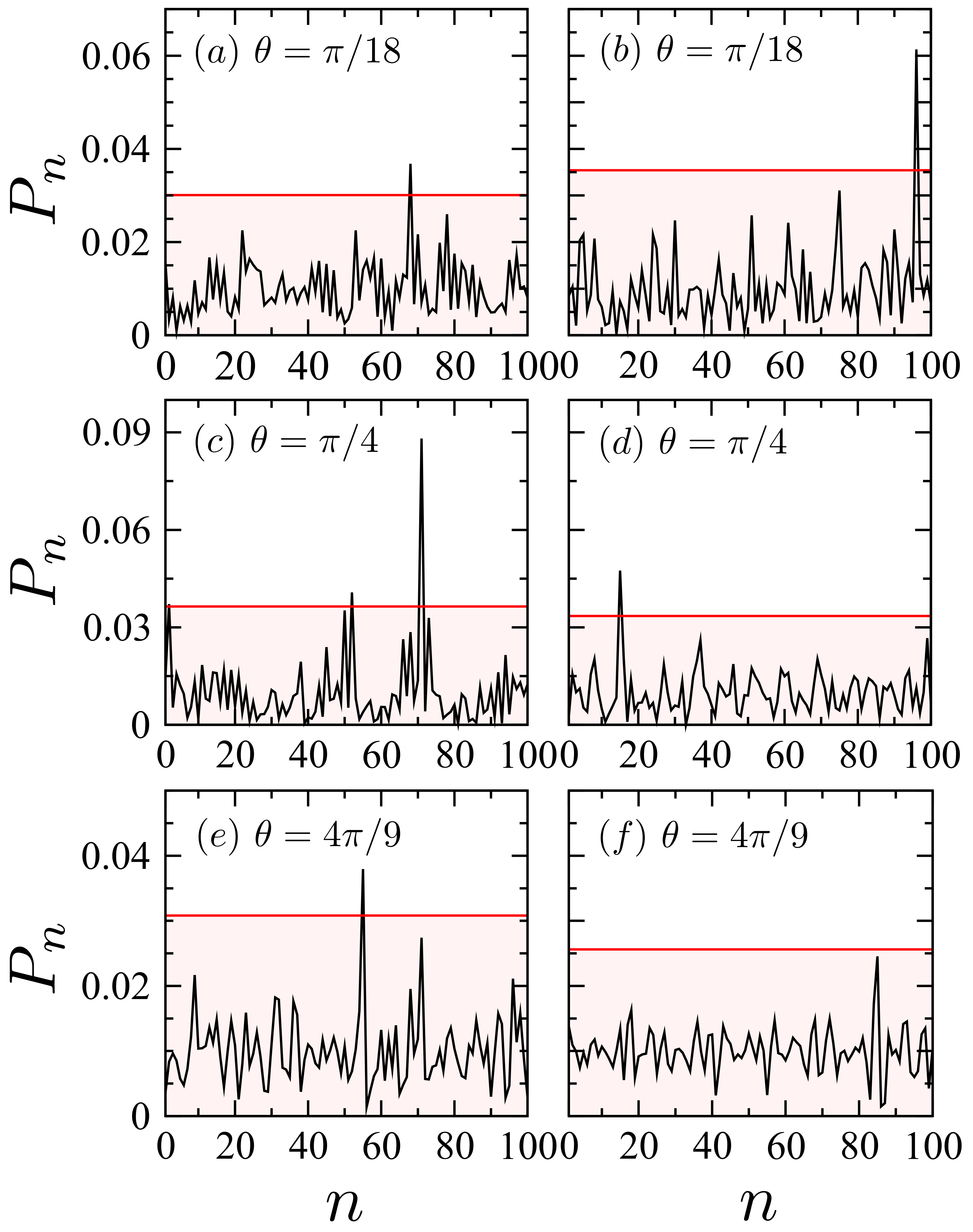}
    \caption{Snapshots of the probability density $P_{n}$ of the quantum walker in a chain with $N=100$ sites after $t=100N$ time steps, for three representative quantum coins: (a)-(b) $\theta=\pi/18$, (c)-(d) $\theta=\pi/4$ (Hadamard), and (e)-(f) $\theta=4\pi/9$.
    Two degrees of disorder are considered: weak disorder in the left column ($W=0.1$) and strong disrorder ($W=0.5$) in the right column. The red horizontal line depicts the probability threshold value  $P_{th}=2\overline{P}_{1/3}$ for the occurrence of RW events.
    }
    \label{figure1}
\end{figure}


\section{Results} 
	
Our results were obtained by following the time evolution of a qubit with initial wave function evenly distributed across all sites of the chain, 
    \begin{eqnarray}
    \label{eq:initial_state}
        \ket{\Psi_{0}}=\frac{1}{\sqrt{2N}}\sum_{n=1}^{N}(\ket{\uparrow}+i\ket{\downarrow})\otimes\ket{n}.
    \end{eqnarray}
We note that the choice of a completely delocalized initial state avoids ambiguity between RWs and the Anderson localization phenomenon, which would likely arise if~the walker started with a initially localized wave function, so allowing just a few modes to act in the evolution of the wave packet and leading to narrow periodic beats over~time.

We begin our discussions by examining the time evolution of the probability density ${P_{n}}$ of the quantum walker as a function of the position~$n$ on a chain with~${N=100}$~sites, over a time period of~$t=100N$. 
Here we define RWs associated with quantum state configurations exhibiting probability amplitudes at some site~$n$ greater than twice the average probability of one-third of the largest amplitudes \cite{PhysRevE.102.052219,rivas2020rogue,PhysRevA.107.012425}, i.e., for $P_n$ above the threshold probability amplitude  $P_{\text{th}}=2\overline{P}_{1/3}$, represented by the red horizontal line in Fig.~\ref{figure1}. 

To understand the influence of different quantum coins on the emergence of RWs in DTQWs, we present in Fig.~\ref{figure1} snapshots of $P_n$ at times when RWs occur. 
We consider three representative configurations of quantum coins: $\theta=\pi/18$ in Figs.~\ref{figure1}(a)-(b); $\theta=\pi/4$ (Hadamard) in Figs.~\ref{figure1}(c)-(d); and $\theta=4\pi/9$ in Figs.~\ref{figure1}(e)-(f), under two distinct disorder situations.
The right panel in Fig.~\ref{figure1} represents quantum walks with weak disorder for~$W=0.1$, while the left one displays results for the strong disorder regime, $W=0.5$. 
We notice that when disorder is weak RWs arise for all configurations of quantum coins. 
However, for strong disorder RWs are not present for quantum coins close to the $\theta = \pi/2$ Pauli-X choice, such as $\theta=4\pi/9$, as indicated in Fig.~\ref{figure1}(f).

 	\begin{figure}[!t]
        \centering
\includegraphics[width=\linewidth]{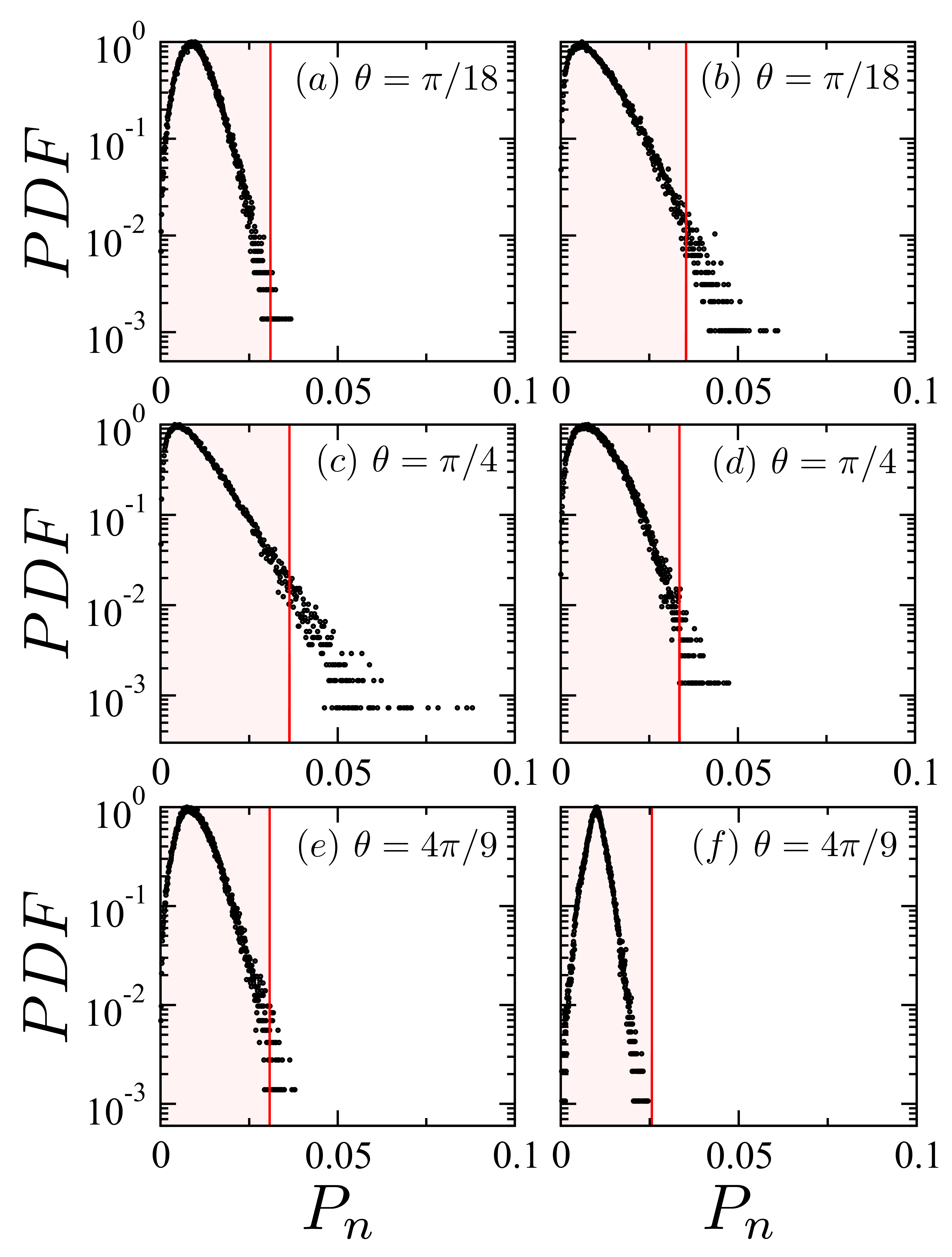}
            \caption{Probability density functions (PDF) of $P_n$ values for the same parameters of  Fig.~\ref{figure1}. 
            The vertical red line marks the threshold $P_{th}=2\overline{P}_{1/3}$ for the RW occurrence. 
            More pronounced non-Gaussian PDFs occur close to the $\theta = \pi/4$ Hadamard choice, while results for $\theta=4\pi/9$ near the Pauli-X choice display Gaussian-like shape.}
        \label{figure2}
    \end{figure}

For a statistical viewpoint, we display in Fig.~\ref{figure2} the probability density function (PDF) of values of $P_n$ for the same configurations shown in Fig.~\ref{figure1}. 
In fact, looking into these PDFs is relevant because they may exhibit another important signature of the occurrence of RWs, which is a non-Gaussian L-shape-like statistics \cite{gao2020optical,PhysRevE.102.052219}. 
The threshold amplitude value $P_{\text{th}}$ is shown in red vertical line in Fig.~\ref{figure2}.
We notice that, consistent with the results in Fig.~\ref{figure1}, all cases exhibit RW events with ${P_{n}}$ surpassing the threshold limit, except for the quantum walk with coin parameter $\theta=4\pi/9$ near Pauli-X in the strongly disordered regime, which shows a Gaussian-like profile. 
In Fig.~\ref{figure2}(c) the Hadamard quantum walk in the weakly disordered regime ($W=0.1$) displays the PDF with largest number of RW events among all cases shown, thus suggesting that this regime favors the emergence of RWs.

    \begin{figure}[!t]
		\centering
		\includegraphics[width=\linewidth]{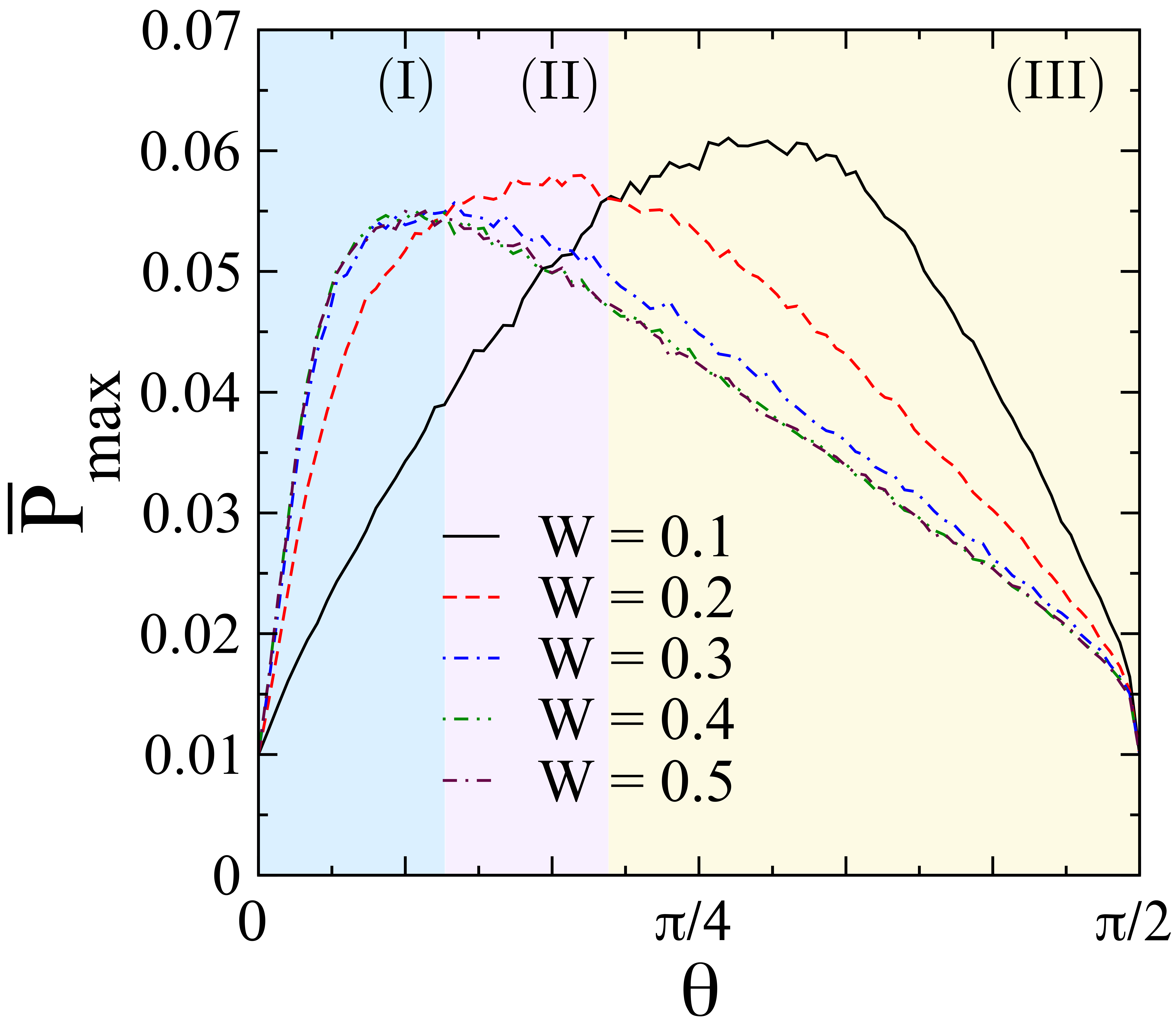}
		\caption{Maximum probability amplitudes of RWs for quantum coins in the range $\theta \in [0,\pi/2]$ and five disorder strengths, $W=0.1, 0.2, 0.3, 0.4, 0.5$, averaged over $10^4$ walk  realizations, in a chain with $N=100$ sites after $t=100N$ time steps. 
Three regions are identified. 
(I) For coins close to $\theta = 0$ Pauli-Z, wavefunction amplitudes are higher for stronger degrees of disorder. 
(II) For coins with intermediate $\theta$ above Pauli-Z and below $\theta = \pi/4$ Hadamard coins,  a moderate disorder strength $(W=0.2)$ yields larger amplitudes. 
(III) Larger $\theta$-values towards the $\theta = \pi/2$ Pauli-X choice lead weakly disordered systems to consistently exhibit higher RW amplitudes.}
		\label{figure3}
    \end{figure}

The results shown in Figs.~\ref{figure1} and~\ref{figure2} evidence that the emergence of RWs in DTQWs depends strongly on the interplay between quantum coins and the degree of disorder. 
Indeed, the amplitude of these rare and unpredictable extreme events varies according to the specific quantum coin applied to the dynamics of the quantum walker. 
In Fig.~\ref{figure3} we present the maximum probability amplitude~$\overline{P}_{\text{max}}$ averaged over $10^4$ independent walk realizations, in a chain with $N=100$ sites after $t=100N$ time steps, for quantum coins in the whole range $\theta \in [0,\pi/2]$ and considering five disorder strengths in the interval $W \in [0.1, 0.5]$. 
We identify in Fig.~\ref{figure3} three regions based on the combination of quantum coins and disorder degree, as follows.  
(I) For coins from the $\theta = 0$ Pauli-Z choice and up to $\theta \approx 0.3$, the average amplitude of RWs tends to be higher for stronger degrees of disorder ($W=0.5$ in Fig.~\ref{figure3}). 
In this case the interaction with coins near Pauli-Z amplifies the effect of disorder, leading to more pronounced RW events. 
(II) In this region, quantum walks with intermediate disorder $(W=0.2)$ exhibit larger probability amplitudes on average. 
This suggests that the occurrence of RWs in this regime is more likely for quantum coins away from both Pauli-Z and Pauli-X choices, $0.3 \lesssim \theta \lesssim 0.6$, and for moderate disorder strengths. 
(III)~Finally, in the regime of weakly disordered quantum walks $(W=0.1)$, the average maximum probability amplitude of RWs remains consistently higher when compared to other disorder degrees. 
This result indicates that for coins with $\theta \gtrsim 0.6$, even for relatively low disorder, there is a higher probability of observing RW events. 
%
%

  \begin{figure}[!t]
		\centering
\includegraphics[width=0.9\linewidth]{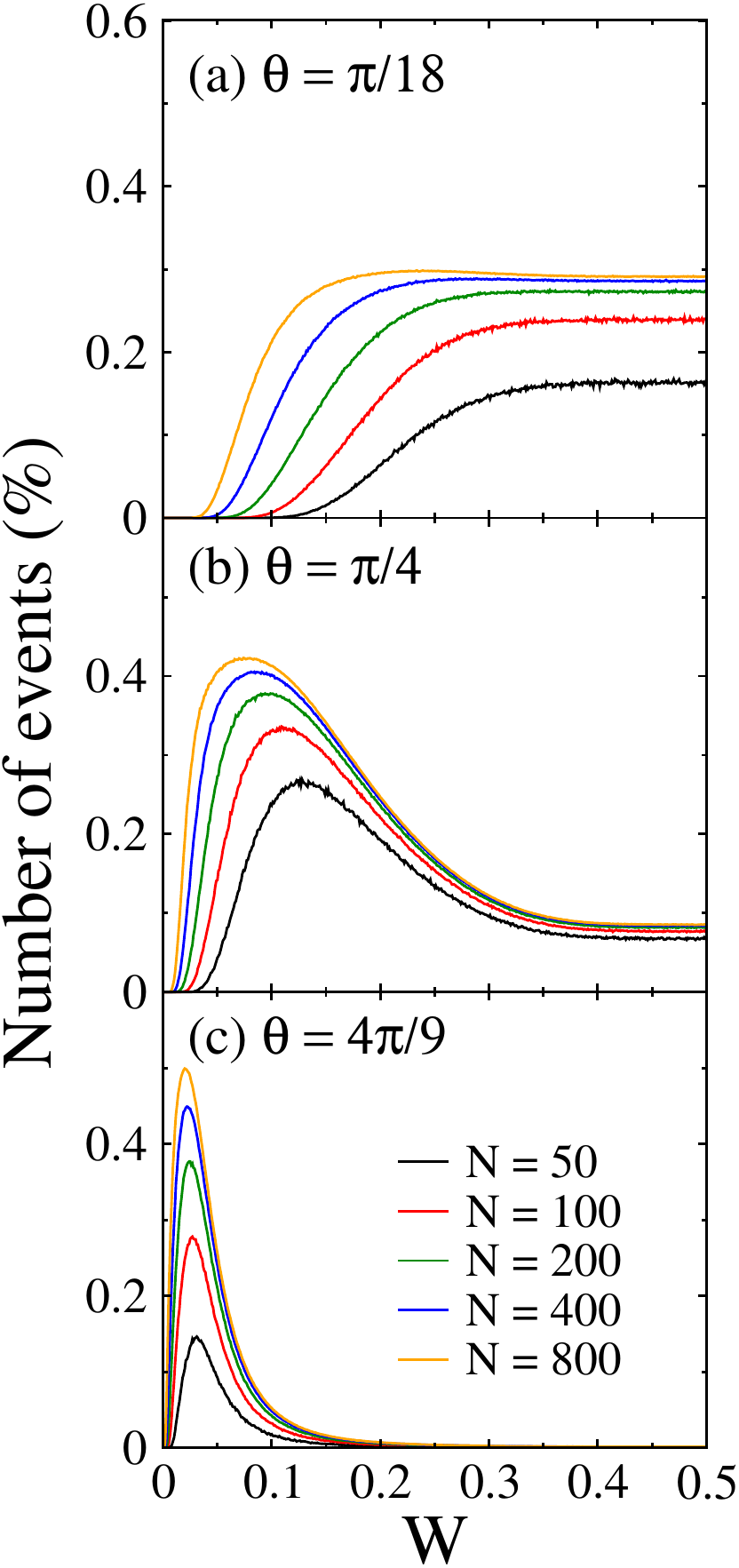}
		\caption{Relative fraction of RW events as a function of the disorder strength~$W$ for $N=50, 100, 200, 400, 800$ chain sizes, with three representative choices of quantum coins: (a)~${\theta=\pi/18}$, (b)~${\theta=\pi/4}$ (Hadamard), and (c)~${\theta=4\pi/9}$. 
  Averages were taken over $10^4$~quantum walk realizations after $t=100N$ time steps.}
		\label{figure4}
	\end{figure}

    \begin{figure}[!t]
		\centering
		\includegraphics[width=\linewidth]{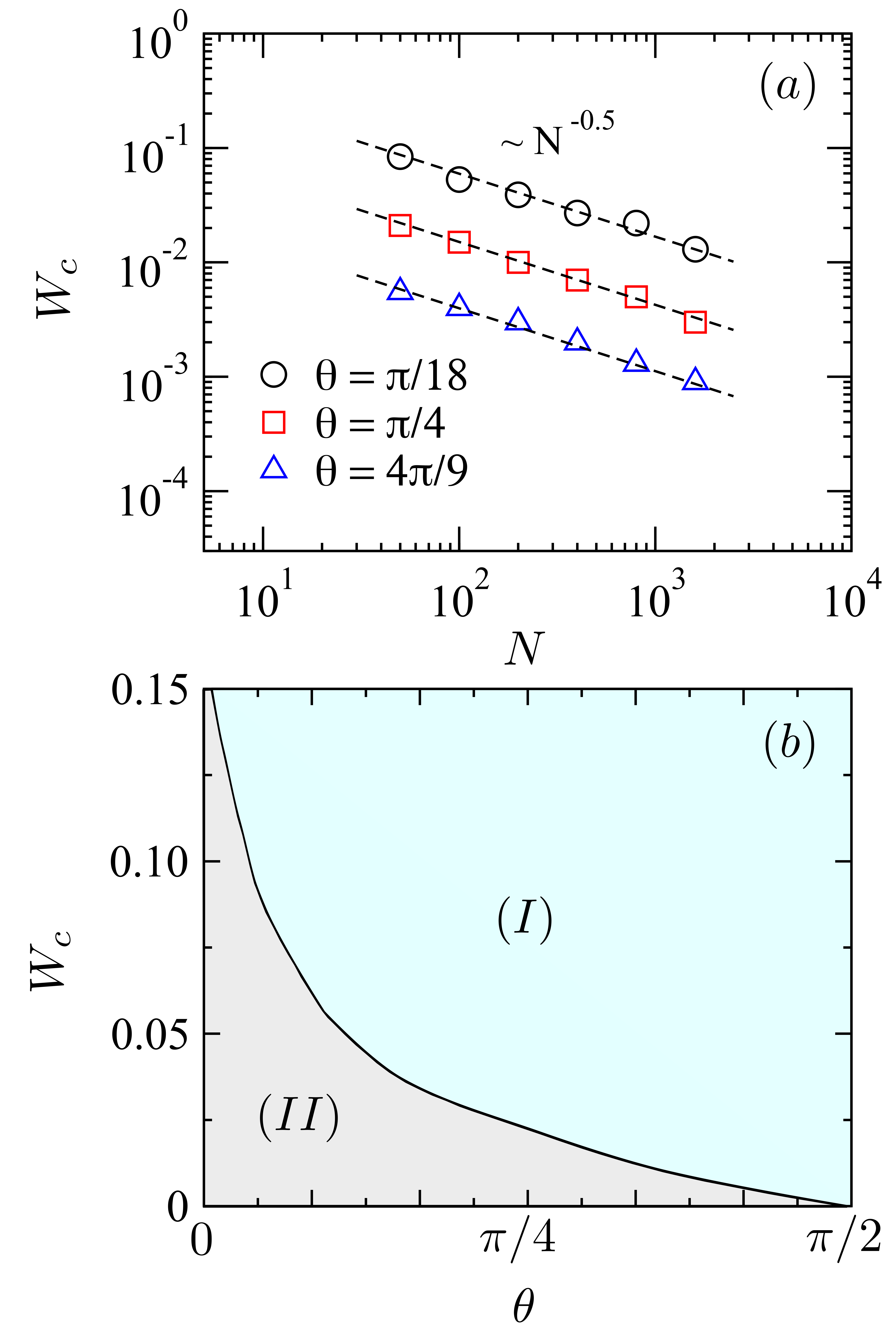}
	\caption{(a) Disorder strength $W_{c}$ above which RWs emerge for three choices of quantum coins: ${\theta=\pi/18}$ (black circles), ${\theta=\pi/4}$ (red squares, Hadamard), and ${\theta=4\pi/9}$ (blue triangles). 
 The scaling behavior $W_{c}\propto N^{-1/2}$ unveils that at $W=W_{c}$ the localization length $\lambda\propto 1/W^{2}$ is of the order of the chain size~$N$, independently of the quantum coin.
 (b)~Colors blue and gray depict regions~(I) and~(II) respectively associated with the presence or not of RW events. 
 }	
		\label{figure5}
    \end{figure}

    \begin{figure}[!t]
        \centering
        \includegraphics[width=\linewidth]{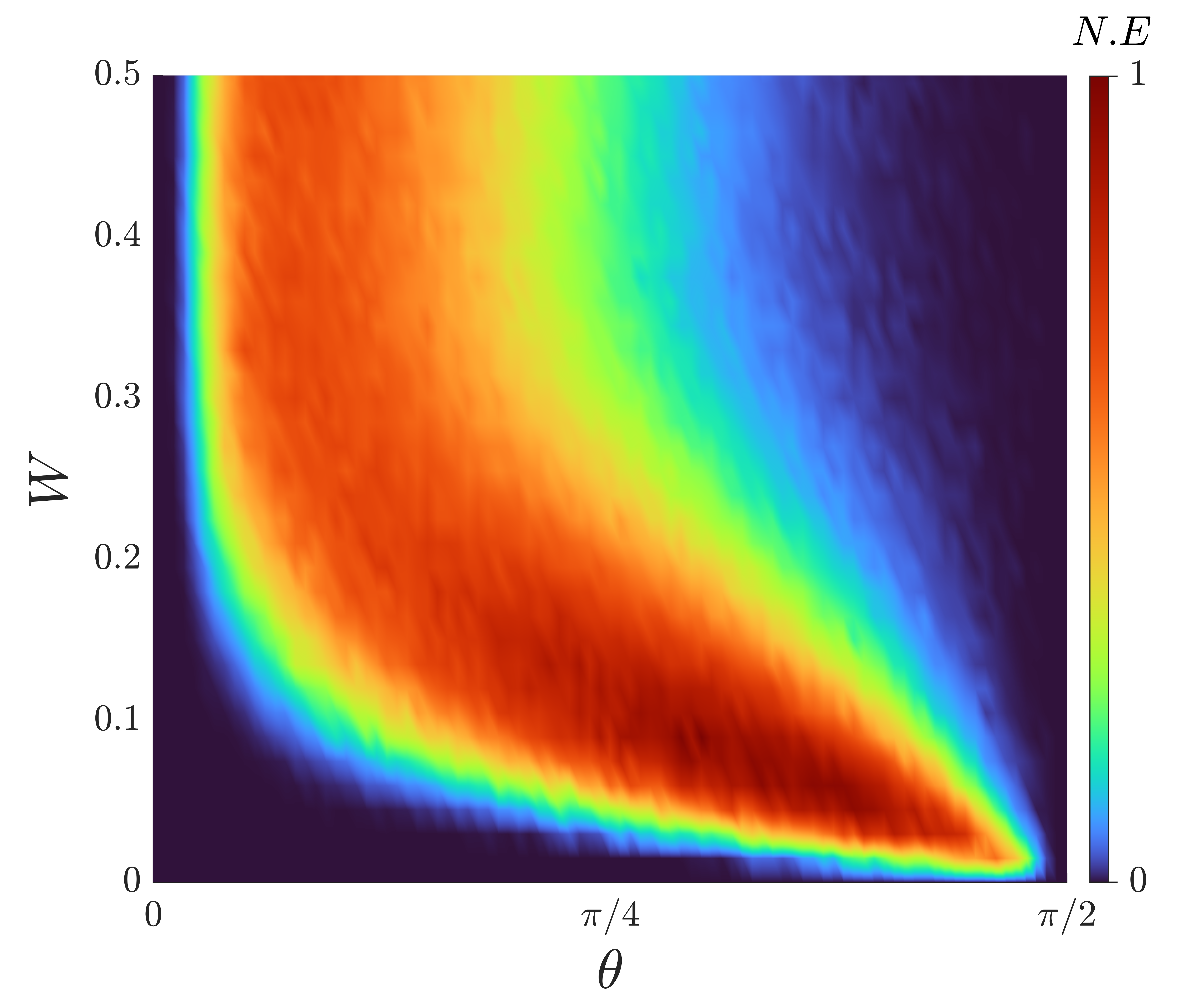}
	\caption{Heatmap of the fraction of RW events in the parameter space $W\in[0,0.5]$ and $\theta\in[0,\pi/2]$, for a chain with $N=100$ sites and $t=100N$. 
 One identifies regions in which the occurrence of RW events is very rare, such as near the points $(\theta, W) = (0, 0)$ and $(\theta, W) = (\pi/2, 0.5)$, as well as configurations of quantum coins and disorder strengths that promote the likely emergence of these events, as in the region $\pi/4 \lesssim \theta \lesssim \pi/2$ and $0 \lesssim W \lesssim 0.2$.}		
		\label{figure6}
    \end{figure}

We now turn to the analysis of the relative fraction of RW events occurring for disorder strengths in the range $W \in [0,0.5]$. 
Figure~\ref{figure4} portraits this quantity averaged over $10^4$ walks after $t=100N$ time steps, for chains with $N=50, 100, 200, 400$, and $800$ sites, showing how the system size influences the statistics and localization behavior of RWs. 
We consider in Fig.~\ref{figure4}(a) the  quantum coin with $\theta=\pi/18$. 
In this case, we observe that the minimum value of~$W$ in the weak disorder regime for the emergence of RWs changes with the system size, despite the high mobility of the quantum walker. 
The number of RW events always saturates on average at relative values no large than~0.3 for all chain sizes and sufficiently large disorder, a result directly related to the localization wavelength of the walker's wavefunction for quantum coins close to Pauli-Z. 
So, before reaching saturation the evolution of the wavepacket is generally characterized by sparse low-amplitude waves that hardly add up to produce rogue events. 

On the other hand, for the Hadamard quantum walk with $\theta=\pi/4$, Fig.~\ref{figure4}(b) shows that the minimum degree of disorder required for the emergence of RWs is reduced. 
In the strongly disordered regime the relative fraction of RW events is much smaller than the corresponding one for $\theta=\pi/18$, Fig.~\ref{figure4}(a). 
In addition, for quantum walks with coins tending to Pauli-X, $\theta=4\pi/9$ in Fig.~\ref{figure4}(c), no RWs arise for disorder strengths $W \gtrsim 0.2$. 
In this regime, localization effects on the wave packet become more pronounced and the walker presents low degree of mobility.

In order to deepen the understanding of these findings, we plot in Fig.~\ref{figure5}(a) the minumum disorder strength~$W_{c}$ above which RWs can emerge as a function of the chain size $N$, for the three previous values of $\theta$. 
Regardless the coin choice, we notice that $W_{c}\propto N^{-1/2}$, so that an increase in the chain size renders the quantum system to be more susceptible to RW events. 
This scaling behavior relates to the fact that the emergence of RWs in this context is induced by disorder, and so these events take place when the associated Anderson localization length~$\lambda$ decreases to a value smaller than the system size, ${\lambda < N}$. 
In fact, in the regime of weak disorder the typical localization length of the eigenstates in quantum walks subjected to random phase shifts exhibits \cite{derevyanko2018anderson} a quadratic dependence on the inverse of the square disorder width, namely, $\lambda=k/W^{2}$, where $k$ is a constant. 
Hence the condition for the emergence of RWs becomes $k/W^{2}<N$, or~$W>(k/N)^{1/2}$, in agreement with Fig.~\ref{figure5}(a). 
Figure~\ref{figure5}(b) displays the dependence of $W_{c}$ on the quantum coin value $\theta$, for a chain with $N=100$ sites and~$t=100N$. 
Colors blue and gray depict regions~(I) and~(II) respectively associated with the presence or not of RW events. 
We note that quantum coins near the $\theta = 0$ Pauli-Z choice are more effective in mitigating the occurrence of~RWs. 
On the other hand, as one considers coins closer to $\theta=\pi/2$ Pauli-X, a monotonic decay of $W_{c}$ becomes apparent, underscoring the substantial influence of quantum coins on the occurrence of RW events in DTQWs. 

Finally, all the above findings can be summarized in the comprehensive mapping shown in Fig.~\ref{figure6} of the relative fraction of RW events in the parameter space defined by the quantum coin parameter ($\theta\in[0,\pi/2]$) and disorder strength ($W\in[0,0.5]$), for a chain with $N=100$ sites and $t=10N$.
We first notice the absence of RWs in the weakly disordered regime, ${0 \lesssim W \lesssim 0.2}$, for quantum coins close to $\theta = 0$ Pauli-Z. 
Indeed, regardless the disorder degree the emergence of RWs is very rare in this regime. 
On the other hand, for quantum coins in the range $\pi/4 \lesssim \theta \lesssim \pi/2$, disorder strengths $0 \lesssim W \lesssim 0.2$ lead to the most likely occurrence of RW events, a trend that fades away as the $\theta=\pi/2$ Pauli-X coin is approached. 
At last, in the intermediate and high disorder regimes, ${0.2 \lesssim W \lesssim 0.5}$, the relative fraction of RW events remains nearly constant for quantum coins $0 \lesssim \theta \lesssim 2\pi/9$, and then starts to decline until entering a less likely region, \linebreak consistent with our previous results.  

\section{Final remarks and Conclusions}
 
In this work, we have delved into the intriguing phenomenon of rogue waves (RWs) within the discrete-time quantum walks (DTQW) protocol. 
Through a comprehensive analysis of different quantum coin configurations and degrees of disorder, we have shed light on the general properties of these rare and unpredictable events in DTQW chains. 
In this context, our investigation revealed a rich interplay between quantum dynamics, disorder, and coin parameters to the emergence of RW events. 

Notably, we have identified distinct regimes where RWs are more likely to manifest, influenced by factors such as quantum phase flucutations, disorder-induced localization, and spatial scaling. 
The transition from weak to strong disorder highlighted the evolution of the RW behavior, with certain coin configurations amplifying or dampening their occurrence. 

The identification of specific parameter ranges of disorder strength and quantum coins in which RWs are exceptionally rare or abundant contributes to the understanding of their controllability, with possible applications in various fields. 
In general lines, our study provides a comprehensive framework for exploring and manipulating RWs in DTQWs. 
The insights gained here not only enrich the overall knowledge of RWs but also offer potential avenues for harnessing their unique properties in future advances.
%


\section{Acknowledgments}

This work was partially supported by the Brazilian agencies CNPq (Conselho Nacional de Desenvolvimento Científico e Tecnológico) and FACEPE (Fundação de Amparo a Ciência e Tecnologia do Estado de Pernambuco).
	
\bibliography{references.bib}

\end{document}